\documentclass[
reprint,
superscriptaddress,
nofootinbib,
amsmath,amssymb,
aps,
]{revtex4-2}
\usepackage{graphicx}
\usepackage{hyperref}
\usepackage{dcolumn}
\usepackage{bm}
\usepackage[mathlines]{lineno}
\usepackage{verbatim}
\begin{document}
\preprint{APS/123-QED}
\title{Nuclear matrix elements for $\lambda$ mechanism of $0\nu\beta\beta$ of $^{48}$Ca in nuclear shell-model: Closure versus nonclosure approach}
\author{Shahariar Sarkar}
\email{shahariar.sarkar@iitrpr.ac.in}
\affiliation{Indian Institute of Technology Ropar, Rupnagar, Punjab-140001, India}
\author{Y. Iwata}
\affiliation{Faculty of Chemistry, Materials and Bioengineering, Kansai University, Japan}
\author{P.K. Raina}
\affiliation{Indian Institute of Technology Ropar, Rupnagar, Punjab-140001, India}
\date{\today}
\begin{abstract}
The $\lambda$ and $m_{\beta\beta}$ mechanisms of neutrinoless double beta decay ($0\nu\beta\beta$) occur with light neutrino exchange via $W_L-W_R$, and $W_L-W_L$ mediation, respectively. In the present study, we calculate the nuclear matrix elements (NMEs) for the $m_{\beta\beta}$ and $\lambda$ mechanisms of $0\nu\beta\beta$, which has origin in the left-right symmetric model with right-handed gauge boson at TeV scale. The NMEs are calculated for one of the $0\nu\beta\beta$ decaying isotope $^{48}$Ca in the interacting nuclear shell-model using the GXPF1A effective interaction of $pf$-shell. The NMEs are calculated in both closure and nonclosure approaches using four different methods: closure, running closure, running nonclosure, and mixed methods. All the NMEs are calculated incorporating the effects of the finite size of nucleons and the revisited higher order terms such as isoscalar and weak magnetism terms of the nucleon currents. Inclusion of the short-range nature of nucleon-nucleon interaction in Miller-Spencer, CD-Bonn, and AV18 parametrizations is also taken care of. The comparative dependence of the running closure and running nonclosure NMEs with the spin-parity of the allowed states of intermediate nucleus $^{48}$Sc, the coupled spin-parity of the two initial decaying neutrons and the final two protons, the cutoff excitation energy of $^{48}$Sc, the cutoff number of states of $^{48}$Sc are also examined. Results show that there are about 2-20\% enhancements in different types of total NMEs, calculated in the nonclosure approach as compared to the closure approach. The significant enhancements are found in the $M_{qGT}$ and $M_{qT}$ type NMEs for the inclusion of the higher-order terms of the nucleon currents.
\end{abstract}

\maketitle
\section{\label{sec:level1}Introduction}
The neutrinoless double beta decay ($0\nu\beta\beta$) is an important weak nuclear process which occurs when two neutrons inside some even-even nuclei converted into two protons and two electrons. In this process neutrino comes as a virtual intermediate particle, thus, it violates lepton number by two units. If this rare process is observed, one can confirm that neutrinos are Majorana particle \cite{PhysRevD.25.2951} which is favoured by many theoretical particle physics models to explain the smallness of neutrino mass \cite{deppisch2012neutrinoless,PhysRevD.25.2951,rodejohann2011neutrino}. This process can also give some ideas about the absolute mass scale of neutrinos which are still unknown \cite{tomoda1991double, RevModPhys.80.481}. Various decay mechanisms such as light neutrino-exchange mechanism \cite{rodin2006assessment,PhysRevC.60.055502}, heavy neutrino-exchange mechanism \cite{vergados2012theory}, left-right symmetric mechanism \cite{PhysRevLett.44.912,PhysRevLett.47.1713}, and the supersymmetric particles exchange mechanism \cite{PhysRevD.34.3457,vergados1987neutrinoless} have been proposed for $0\nu\beta\beta$.

In the present work, we focus on the $\lambda$ mechanism ($W_L$-$W_R$ exchange) along with the standard $m_{\beta\beta}$ mechanism ($W_L-W_L$ exchange) of the $0\nu\beta\beta$ mediated by light neutrinos \cite{vsimkovic2017lambda}. The $\lambda$ mechanism has origin in the left-right symmetric mechanism with right-handed gauge boson at the TeV scale \cite{vsimkovic2017lambda}. 
The decay rate in all of these mechanisms is related to the nuclear matrix elements (NMEs) and absolute neutrino mass. These NMEs are calculated using theoretical nuclear many-body models \cite{engel2017status} such as the quasiparticle random phase approximation (QRPA) \cite{vergados2012theory}, the interacting shell-model (ISM) \cite{PhysRevLett.100.052503,PhysRevC.81.024321,PhysRevC.88.064312,PhysRevLett.113.262501,PhysRevLett.116.112502}, the interacting boson model (IBM) \cite{PhysRevC.79.044301,PhysRevLett.109.042501}, the generator coordinate method (GCM) \cite{PhysRevLett.105.252503}, the energy density functional (EDF) theory \cite{PhysRevLett.105.252503,PhysRevC.90.054309} and the projected Hartree-Fock Bogolibov model (PHFB) \cite{PhysRevC.82.064310}. 

Recently in Ref.~\cite{PhysRevC.92.055502}, the revisited formalism for the light neutrino-exchange mechanism of $0\nu\beta\beta$ was exploited to include the effects of isoscalar term of nucleon currents. Using the revisited formalism of Ref.~\cite{PhysRevC.92.055502}, the NMEs for $\lambda$, and $m_{\beta\beta}$  mechanisms of $0\nu\beta\beta$ were calculated using the QRPA model for several $0\nu\beta\beta$ decaying isotopes using closure approximation in Ref.~\cite{vsimkovic2017lambda}. In this case, the weak magnetism term of the nucleon currents was also considered for calculating the NMEs of $m_{\beta\beta}$ mechanism. Most of the NMEs relevant for $\lambda$, and $m_{\beta\beta}$ mechanisms were calculated using ISM in Ref. \cite{PhysRevC.98.035502} using closure approximation for $^{48}$Ca and few other $0\nu\beta\beta$ decaying isotopes. In this case, some of the NMEs were calculated without including the higher-order terms (isoscalar and weak magnetism) of the nucleon currents. 
\begin{figure*}
\centering
\includegraphics[trim=0cm 1cm 0cm 0cm,width=\linewidth]{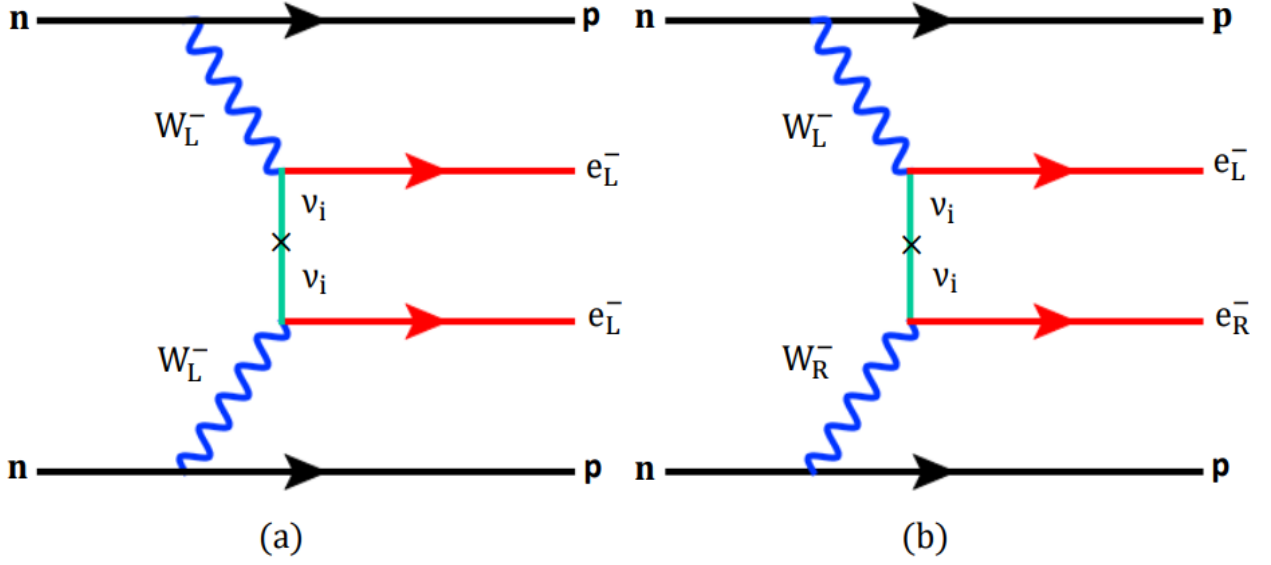}
\caption{\label{fig:feynman}(Color online) The Feynman diagrams for $0\nu\beta\beta$ via (a) $W_L-W_L$ mediation ($m_{\beta\beta}$ mechanism) and (b) $W_L-W_R$ mediation ($\lambda$ mechanism) with light neutrinos exchange.}
\end{figure*}

In the present study,  the NMEs for the $\lambda$, and  $m_{\beta\beta}$ mechanisms of $0\nu\beta\beta$ of $^{48}$Ca are calculated using ISM incorporating the effects of isoscalar and weak magnetism terms of nucleon currents. Earlier, nonclosure approach was applied in ISM for calculating NMEs only for the $m_{\beta\beta}$ mechanism of $0\nu\beta\beta$ of $^{48}$Ca in Ref. \cite{PhysRevC.88.064312}. In this work, we use both closure and nonclosure approaches to calculate the NMEs for both the $\lambda$, and  $m_{\beta\beta}$ mechanisms of $0\nu\beta\beta$ of $^{48}$Ca.
The $0\nu\beta\beta$ process for $^{48}$Ca is written as \begin{equation}
    ^{48}\text{Ca}\rightarrow^{48}\text{Ti}+e^-+e^-.
\end{equation}

In our calculation, the short range-nature of the nucleon-nucleon interaction was also taken care of in Miller-Spencer, CD-Bonn, and AV18 type short-range correlations (SRC) parametrization \cite{PhysRevC.60.055502,PhysRevC.81.024321}. The NMEs are calculated using widely used $pf$-shell interaction GXPF1A. Using both the closure and nonclosure approaches, the NMEs are calculated in four different methods: closure, running closure, running nonclosure, and mixed methods \cite{PhysRevC.88.064312}.

This paper is organized as follows. In section II, the expression for decay rate and the theoretical formalism to calculate NMEs for the $\lambda$ and $m_{\beta\beta}$ mechanisms of $0\nu\beta\beta$ are presented. The descriptions of different methods of NME calculation based on closure and nonclosure approaches are given in section III. The results and discussion are presented in section IV. The summary of the work is given in section V.
\section{Decay rate For $\lambda$ mechanism of $0\nu\beta\beta$}
The Feynman diagrams for the light neutrino-exchange $m_{\beta\beta}$, and $\lambda$ mechanisms are shown in Fig. \ref{fig:feynman}. The decay rate for $0\nu\beta\beta$, when both the mechanisms coexist, can be written as \cite{vsimkovic2017lambda,PhysRevC.92.055502}
\begin{equation}
 [T_{1/2}^{0\nu}]^{-1}=\eta_{\nu}^{2}C_{mm}+\eta_{\lambda}^{2}C_{\lambda\lambda}+\eta_{\nu}\eta_{\lambda}\text{cos}~\psi C_{m\lambda},
\end{equation}
where $\eta_{\nu}$ is effective lepton number violating parameters for $W_L-W_L$ exchange and $\eta_{\lambda}$ is effective lepton number violating parameters for $W_L-W_R$ exchange, which are given by \cite{vsimkovic2017lambda} 
\begin{align}
\eta_{\nu} & =\frac{m_{\beta\beta}}{m_{e}},\quad\eta_{\lambda}=\lambda|\sum_{j=1}^{3}m_{j}U_{ej}T_{ej}^{*}|,\\
\psi & =\textrm{arg}\left[\left(\sum_{j=1}^{3}m_{j}U_{ej}^{2}\right)\left(\sum_{j=1}^{3}U_{ej}T_{ej}^{*}\right)\right].
\end{align}
Here, $U$, and $T$ are the $3\times 3$ block matrices in flavor space, which constitute a generalization of the Pontecorvo-Maki-Nakagawa-Sakata matrix, namely the 6 $\times$ 6 unitary neutrino mixing matrix \cite{vsimkovic2017lambda,PhysRevC.92.055502}.
The coefficients $C_I$ ($I$ = $mm$, $m\lambda$ and $\lambda\lambda$) are linear combinations of products of nuclear matrix elements and phase-space factors \cite{vsimkovic2017lambda}
\begin{align}
C_{mm} & =g_{A}^{4}M_{\nu}^{2}G_{01},\\
C_{m\lambda} & =-g_{A}^{4}M_{\nu}(M_{2-}G_{03}-M_{1+}G_{04)},\\
C_{\lambda\lambda} & =g_{A}^{4}(M_{2-}^{2}G_{02}+\frac{1}{9}M_{1+}^{2}G_{011}-\frac{2}{9}M_{1+}M_{2-}G_{010}).
\end{align}
Calculated values of phase-space factors $G_{0i} (i = 1, 2, 3, 4, 10 ~\text{and}~ 11)$ for different  $0\nu\beta\beta$ decaying nuclei are given in Ref. \cite{PhysRevC.92.055502}.
Matrix elements required in the expression of $C_I$ are \cite{vsimkovic2017lambda}
\begin{align}
M_{\nu} & =M_{GT}-\frac{M_{F}}{g_{A}^{2}}+M_{T},\\
M_{\nu\omega} & =M_{\omega GT}-\frac{M_{\omega F}}{g_{A}^{2}}+M_{\omega T},\\
M_{1+} & =M_{qGT}+3\frac{M_{qF}}{g_{A}^{2}}-6M_{qT},\\
M_{2-} & =M_{\nu\omega}-\frac{1}{9}M_{1+}.
\end{align}
The ($M_{GT,\omega GT,qGT}$), ($M_{F,\omega F,qF}$), and ($M_{T,\omega T,qT}$) matrix elements of the scalar two-body transition operator $\mathcal{O}_{12}^\alpha$  of $0\nu\beta\beta$
can be expressed as \cite{PhysRevLett.113.262501}
\begin{eqnarray}
\label{Eq:NMEMAIN}
&&M^{0\nu}_{\alpha}=\langle f|\tau_{-1}\tau_{-2}\mathcal{O}_{12}^\alpha|i\rangle
\end{eqnarray}
where, $|i\rangle$, and $|f\rangle$ are the initial and the final $0^+$ ground state (g.s) for $0\nu\beta\beta$ decay, respectively, $\alpha={(GT,F,T,\omega GT,\omega F,\omega T,qGT,qF,qT)}$, $\tau_{-}$ is the isospin annihilation operator. The scalar two-particle transition operators $\mathcal{O}_{12}^\alpha$ of $0\nu\beta\beta$ containing spin and  radial neutrino potential operators can be written as
\begin{eqnarray}
\label{eq:opnc}
\mathcal{O}_{12}^{GT,\omega GT,qGT}&&=\tau_{1-}\tau_{2-}(\mathbf{\sigma_1.\sigma_2)}H_{GT,\omega GT,qGT}(r,E_k),
\nonumber\\
\mathcal{O}_{12}^{F,\omega F,qF}&&=\tau_{1-}\tau_{2-}H_{F,\omega F,qF}(r,E_k),
\\
\mathcal{O}_{12}^{T,\omega T,qT}&&=\tau_{1-}\tau_{2-}S_{12}H_{T,\omega T,qT}(r,E_k),
\nonumber
\end{eqnarray}
where, $S_{12}=3(\mathbf{\sigma_1 .\hat{r})(\sigma_2.\hat{r})-(\sigma_1.\sigma_2)}$, $\mathbf{r=r_1-r_2}$, and $r=|\mathbf{r}|$ is inter nucleon distance of the decaying nucleons.

The neutrino potential for $\lambda$ mechanism of $0\nu\beta\beta$ are given as integral over Majorana neutrino momentum q \cite{PhysRevC.88.064312}:
\begin{equation}
\label{eq:npnc}
H_\alpha (r,E_{k})=\frac{2R}{\pi}\int_{0}^{\infty}\frac{f_\alpha(q,r)qdq}{q+E_{k}-(E_{i}+E_{f})/2}
\end{equation}
where $R$ is the radius of the parent nucleus, $E_k$ is the energy of the intermediate states, $E_i$ is the energy of the initial state, $E_f$ is the energy of the final state, and the $f_\alpha(q,r)$ factor contains the form factors that incorporates the effects of finite nucleon size (FNS), and higher-order currents (HOC) of nucleons \cite{PhysRevC.60.055502}. The $f_\alpha(q,r)$ factor can be written in terms of radial dependence, spherical Bessel function $j_p(qr) (p=0,1,2~\text{and}~3)$, and FNS+HOC coupling form factors as \cite{vsimkovic2017lambda}
\begin{eqnarray}
f_{GT}(q,r) &=&\frac{j_{0}(qr)}{g_{A}^{2}}\left(g_{A}^{2}(q^{2})-\frac{g_{A}(q^{2})g_{P}(q^{2})}{m_{N}}\frac{q^{2}}{3}\right.\nonumber\\
&&\left. +\frac{g_{P}^{2}(q^{2})}{4m_{N}^{2}}\frac{q^{4}}{3}+\left(2\frac{g_{M}^{2}(q^{2})}{4m_{N}^{2}}\frac{q^{2}}{3}\right)\right),\\
f_{F}(q,r)&=&g_{V}^{2}(q^{2})j_{0}(qr),\\
f_{T}(q,r)&=&\frac{j_{2}(qr)}{g_{A}^{2}}\left(\frac{g_{A}(q^{2})g_{P}(q^{2})}{m_{N}}\frac{q^{2}}{3}-\frac{g_{P}^{2}(q^{2})}{4m_{N}^{2}}\frac{q^{4}}{3}\right.\nonumber\\
&&\left.+\frac{g_{M}^{2}(q^{2})}{4m_{N}^{2}}\frac{q^{2}}{3}\right),\\
f_{\omega GT}(q,r)&=&\frac{q}{(q+E_{k}-(E_{i}+E_{f})/2)}f_{GT}(q,r),\\
f_{\omega F}(q,r)&=&\frac{q}{(q+E_{k}-(E_{i}+E_{f})/2)}f_{F}(q,r),\\
f_{\omega T}(q,r)&=&\frac{q}{(q+E_{k}-(E_{i}+E_{f})/2)}f_{T}(q,r),\\
f_{qGT}(q,r)&=&\left( \frac{g_{A}^{2}(q^{2})}{g_{A}^{2}}q+3\frac{g_{P}^{2}(q^{2})}{g_{A}^{2}}\frac{q^{5}}{4m_{N}^{2}}\right. \nonumber\\
&&\left. +\frac{g_{A}(q^{2})g_{P}(q^{2})}{g_{A}^{2}}\frac{q^{3}}{m_{N}}\right) rj_{1}(q,r),\label{Eq:hqgt}\\
f_{qF}(q,r)&=&rg_{V}^{2}(q^{2})j_{1}(qr)q,\\
f_{qT}(q,r)&=&\frac{r}{3}\left(\left(\frac{g_{A}^{2}(q^{2})}{g_{A}^{2}}q-\frac{g_{P}(q^{2})g_{A}(q^{2})}{2g_{A}^{2}}\frac{q^{3}}{m_{N}}\right)j_{1}(qr)\right.\nonumber\\
&&-\left. \left(9\frac{g_{P}^{2}(q^{2})}{2g_{A}^{2}}\frac{q^{5}}{20m_{N}^{2}}\left[ 2j_{1}(qr)/3-j_{3}(qr)\right]\right) \right),\nonumber\\
\end{eqnarray}
where one can write in dipole approximation \cite{PhysRevC.60.055502}
\begin{eqnarray}
g_V(q^2)&=&\frac{g_V}{\left(1+\frac{q^2}{M_V^2}\right)^2},\\
g_A(q^2)&=&\frac{g_A}{\left(1+\frac{q^2}{M_A^2}\right)^2},\\
g_M(q^2)&=&(\mu_p-\mu_n)g_V(q^2),\\
g_{P}(q^{2})&=&\frac{2m_{p}g_{A}(q^{2})}{(q^{2}+m_{\pi}^{2})}\left(1-\frac{m_{\pi}^{2}}{M_{A}^{2}}\right).
\end{eqnarray}
 $\mu_p-\mu_n=4.7, M_V=850~\textrm{MeV}, M_A=1086~\textrm{MeV}$ $m_p$ and $m_\pi$ are the mass of protons and pions \cite{PhysRevC.88.064312}. In the present calculation, vector constant $g_V=1.0$ and bare axial-vector constant $g_A=1.27$ \cite{PhysRevC.101.014307} was used.
Both the pseudo scalar and weak magnetism terms of the nucleon currents are included in $f_{GT,T,\omega GT,\omega T}(q,r)$ factors whereas pseudo scalar term is included in  $f_{qGT,qT}(q,r)$ factors \cite{vsimkovic2017lambda}. 

The short range nature of the two-nucleon interaction is taken care by multiplying relative harmonic oscillator wavefunciton $\psi_{nl}$ in radial integral $\langle n',l'|H_\alpha(r)|n,l\rangle$ with a correlation function $f(r)$ \cite{PhysRevC.81.024321};
\begin{equation}
    \psi_{nl}(r)\longrightarrow[1+f(r)]\psi_{nl}(r),
\end{equation}
where, $f(r)$ can be parametrized as \cite{PhysRevC.79.055501}
\begin{equation}
    f(r)=-ce^{ar^2}(1-br^2).
\end{equation}
The parameters $a$, $b$ and $c$ for Miller-Spencer, CD-Bonn and AV18 type SRC parametrization are given in Ref.\cite{PhysRevC.81.024321}
\section{\label{sec:III}Closure, nonclosure, and mixed methods}
\textit{Closure approach}:
In closure approach, one replaces the term $[E_{k}-(E_{i}+E_{f})/2]$ in Eq.~(\ref{eq:npnc}) by an average closure energy $\langle E\rangle$:
\[[E_{k}-(E_{i}+E_{f})/2]\rightarrow \langle E\rangle.\]
In closure approach, neutrino potential of Eq.~(\ref{eq:npnc}) becomes
\begin{equation}
\label{eq:npc}
H_\alpha (r)=\frac{2R}{\pi}\int_{0}^{\infty}\frac{f_\alpha(q,r)qdq}{q+\langle E \rangle,}
\end{equation}
and the transition operators of $0\nu\beta\beta$ of Eq.~(\ref{eq:opnc}) can be re-written as
\begin{eqnarray}
O_{12}^{GT,\omega GT,qGT}&&=\tau_{1-}\tau_{2-}(\mathbf{\sigma_1.\sigma_2)}H_{GT,\omega GT,qGT}(r),
\nonumber\\
\label{eq:opc}
O_{12}^{F,\omega F,qF}&&=\tau_{1-}\tau_{2-}H_{F,\omega F,qF}(r),
\\
O_{12}^{T,\omega T,qT}&&=\tau_{1-}\tau_{2-}S_{12}H_{T,\omega T,qT}(r),
\nonumber
\end{eqnarray}
Closure approach has significant advantage over nonclosure approach because it eliminates the complexity of calculating NMEs in terms of excitation energy of large number of the intermediate states, which can be computationally challenging for heavy nuclear systems. This approximation is also very good as the values of $q$ that dominate the matrix elements are of the order of $\sim$100-200 MeV, whereas the relevant excitation energies of the intermediate states are about 10 MeV \cite{PhysRevC.88.064312}. One important part of closure approximation is to use a suitable value of average closure energy $\langle E \rangle$ that will take care the combine effects of a large number of intermediate states. In the present work, we have used standard closure energy $\langle E\rangle=7.72$ MeV \cite{PhysRevC.81.024321,PhysRevC.88.064312}.

\textit{Nonclosure approach}:
In nonclosure approach, one calculates the neutrino potential of Eq.~(\ref{eq:npnc}) explicitly in terms of energy $E_k$ of large number of virtual intermediate states $|k\rangle$ (for our case ($^{48}$Sc). For our nonclosure calculations, we have used \cite{PhysRevC.88.064312}
\begin{equation}
    E_{k}-(E_{i}+E_{f})/2\rightarrow 1.9 \text{MeV}+E_k^{*},
\end{equation}
where $E_k^{*}$ is the excitation energy of the intermediate states $|k\rangle$.

Descriptions of four different methods: closure, running closure, running nonclosure, and mixed methods to calculate NMEs of $0\nu\beta\beta$ based on closure and nonclosure approaches are given below.

\textbf{Closure method:}
In closure method, the NMEs are calculated using transition operator of Eq.~(\ref{eq:opc}) for $\lambda$ mechanism of $0\nu\beta\beta$  and the neutrino potential for closure approximation defined in Eq.~(\ref{eq:npc}). In this method NMEs defined in Eq.~(\ref{Eq:NMEMAIN}) can be written as sum over products of two-nucleon transfer amplitudes (TNAs) and anti-symmetric non-reduced two-body matrix elements $\langle k_1',k_2',JT|\tau_{-1}\tau_{-2}{O}_{12}^\alpha|k_1,k_2,JT\rangle_A$ as \cite{PhysRevLett.113.262501}
\begin{eqnarray}
\label{Eq:NMETNA}
&&\mathcal{M}^{0\nu}_{\alpha}=\sum_{m,J,k_1'\leqslant k_2',k_1\leqslant k_2}\text{TNA}(f,m,k_1', k_2', J_m)\nonumber\\
&&\text{TNA}(i,m,k_1, k_2, J_m)\times\langle k_1',k_2':JT|\tau_{-1}\tau_{-2}{O}_{12}^\alpha|k_1,k_2:JT\rangle_A,\nonumber\\
\end{eqnarray}
where $k$ stands for the set of spherical quantum numbers $(n; l; j)$, and A denotes that the two-body matrix elements are obtained using anti-symmetric two-nucleon wavefunction. In our study, $|i\rangle$ is $0^+$ g.s. of the parent nucleus $^{48}$Ca, $|m\rangle$ is the large number of states of intermediate nucleus ($^{46}$Ca) with allowed spin-parity ($J^{\pi}$) (for TNA calculation),  $|f\rangle$ is the $0^+$ g.s. of the granddaughter nucleus $^{48}$Ti, and $k$ has the spherical quantum numbers for $0f_{7/2}$, $0f_{5/2}$, $1p_{3/2}$, and $1p_{1/2}$ orbitals. Complete expression of anti-symmetric non-reduced two-body matrix elements (TBMEs) is given in Refs. \cite{PhysRevC.101.014307,PhysRevC.81.024321}.

TNA is calculated with a large set of intermediate states $|m\rangle$ of the $(n-2)$ nucleons system ($^{46}$Ca in the present study), where $n$ is the number of nucleons for the parent nucleus.
TNA is given by \cite{PhysRevLett.113.262501}
\begin{equation}
  \text{TNA}(f,m,k_1', k_2', J_m)=\frac{\langle f||A^+(k_1', k_2' ,J_{m})||m\rangle}{\sqrt{2J_0+1}}.
\end{equation}
Here, 
\begin{equation}
  A^+(k_1', k_2' ,J_{m})=\frac{[ a^{+}(k_1')\otimes a^{+} (k_2')]^{J_{m}}_{M}}{\sqrt{1+\delta_{{k_1'}{k_2'}}}}
\end{equation}
is the two particle creation operator of rank $J$, $J_m$ is the spin of the allowed states of $^{46}$Ca, $J_0$ is spin of $|i\rangle$ and $|f\rangle$. In Eq. (\ref{Eq:NMETNA}) $J_m$=$J$ when $J_0$=0 \cite{PhysRevLett.113.262501}. 
The TNA is normalized such that
\begin{equation}
\label{eq:tnap^2}
   \text{TNA}^2=n_p(n_p-1)/2
\end{equation}
for the removal of two protons and
\begin{equation}
\label{eq:tnan^2}
\text{TNA}^2=n_n(n_n-1)/2
\end{equation} for the removal of two neutrons, where $n_{p(n)}$ are the total number of protons (neutrons) in the model-space \cite{PhysRevLett.113.262501}.

\textbf{Running closure method:}
In running closure method, one uses the same $0\nu\beta\beta$ transition operator and neutrino potential as closure method. However, in this method one gets the true virtual intermediate nucleus after one neutron from parent nucleus decay into one proton. In the present study $^{48}$Sc is the true virtual intermediate nucleus.  
For convenience we can write the partial nuclear matrix elements of running closure method in proton-neutron (pn) formalism as function of the spin-parity of the states of the intermediate  nucleus ($J_{k}^{\pi}$), the coupled-spin of two decaying protons or the two final created protons ($J^{\pi}$), and the excitation energy of the states of the virtual intermediate nucleus ($E_k^{*}$) as sum over products of one body transition density (OBTD) and non anti-symmetric reduced two-body matrix elements $\langle k_1',k_2':J||\tau_{-1}\tau_{-2}{O_{12}^\alpha}||k_1,k_2:J\rangle$ as \cite{PhysRevC.88.064312,PhysRevC.77.045503}:
\begin{eqnarray}
\label{eq:rcpartial}
&&\mathcal{M}_{\alpha}^{0\nu}(J_{k},J,E_{k}^{*})=\sum_{k'_{1}k'_{2}k_{1}k_{2}}\sqrt{(2J_{k}+1)(2J_{k}+1)(2J+1)}\nonumber\\
&&\times(-1)^{j_{k1}+j_{k2}+J}
\left\{ \begin{array}{ccc}
j_{k1^{'}} & j_{k1} & J_{k}\\
j_{k2} & j_{k2^{'}} & J
\end{array}\right\}\text{OBTD}(k,f,k'_{2},k_{2},J_{k})\nonumber\\ &&\times \text{OBTD}(k,i,k'_{1},k_{1},J_{k})\langle k_1',k_2':J||\tau_{-1}\tau_{-2}{O_{12}^\alpha}||k_1,k_2:J\rangle,\nonumber\\
\label{eq:mjjkrc}
\end{eqnarray}
Here $k_1$ represents set of spherical quantum numbers ($n_1,l_1,j_1$) for an orbital. Now the final NME in running closure method can be written as
\begin{eqnarray}
\mathcal{M}_{\alpha}^{0\nu}(E_c)=\sum_{J_k,J,E_{k}^{*}\leqslant E_c}\mathcal{M}_{\alpha}^{0\nu}(J_{k},J,E_{k}^{*}),
\end{eqnarray}
where $E^{*}_k$ of each allowed $J_{k}^{\pi}$ of intermediate nucleus can run up to cutoff excitation energy $E_{c}$ in the summation. Considering states whose excitation energy $E^{*}_k$ goes up to $E_{c}$ gives almost constant NMEs when $E_{c}$ is large enough.
OBTD in proton-neutron formalism can be written as \cite{PhysRevC.88.064312}
\begin{equation}
\label{eq:obtd}
  \text{OBTD}(k,i,k'_{1},k_{1},\mathcal{J})=\frac{\langle k||[a_{k'_{1}}^{+}\otimes\widetilde{a}_{k_{1}}]_\mathcal{J}||i\rangle}{\sqrt{2\mathcal{J}+1}}, 
\end{equation}
where $a_{k'_{1}}^{+}$ and $\widetilde{a}_{k_{1}}$ are the one particle creation and annihilation operator, respectively. In the present study $|k\rangle$ is the large number of virtual intermediate states of $^{48}$Sc for different allowed $J_k^{\pi}$.

\textbf{Running nonclosure method:}
In running nonclosure method, NMEs are calculated with the nonclosure $0\nu\beta\beta$ transition operators given in Eq.~(\ref{eq:opnc}). Nonclosure neutrino potential defined in Eq.~(\ref{eq:npnc}) are calculated explicitly in terms of excitation energy of large number allowed states of intermediate nucleus ($^{48}$Sc). Partial NMEs for running nonclosure in proton-neutron formalism can be defined as \cite{PhysRevC.88.064312,PhysRevC.77.045503}
\begin{eqnarray}
\label{eq:rncpartial}
&&M_{\alpha}^{0\nu}(J_{k},J,E_{k}^{*})=\sum_{k'_{1}k'_{2}k_{1}k_{2}}\sqrt{(2J_{k}+1)(2J_{k}+1)(2J+1)}\nonumber\\
&&\times(-1)^{j_{k1}+j_{k2}+J}
\left\{ \begin{array}{ccc}
j_{k1^{'}} & j_{k1} & J_{k}\\
j_{k2} & j_{k2^{'}} & J
\end{array}\right\}\text{OBTD}(k,f,k'_{2},k_{2},J_{k})\nonumber\\ &&\times \text{OBTD}(k,i,k'_{1},k_{1},J_{k})\langle k_1',k_2':J||\tau_{-1}\tau_{-2}{\mathcal{O}_{12}^\alpha}||k_1,k_2:J\rangle,\nonumber\\
\label{eq:mjjkrc}
\end{eqnarray}
where the above Eq. (\ref{eq:rncpartial}) is similar to Eq. (\ref{eq:rcpartial}), except the transition operator $\mathcal{O}_{12}^{0\nu}$, which has explicit $E_{k}^{*}$ dependence.
The final NME in running nonclosure method is given by \cite{PhysRevC.88.064312}
\begin{eqnarray}
{M}_{\alpha}^{0\nu}(E_c)=\sum_{J_k,J,E_{k}^{*}\leqslant E_c}{M}_{\alpha}^{0\nu}(J_{k},J,E_{k}^{*}).
\end{eqnarray}
Complete expression of non anti-symmetric reduced TBMEs for running nonclosure and running closure methods is given in Ref. \cite{PhysRevC.88.064312}.

\textbf{Mixed method:}
The mixed method is the superposition of running nonclosure, running closure, and closure methods. NMEs in the mixed method are written as \cite{PhysRevC.88.064312}
\begin{equation}
\label{eq:nmemixed}
   \bar{M}_{\alpha}^{0\nu}(E_c)=M_{\alpha}^{0\nu}(E_c)-\mathcal{M}_{\alpha}^{0\nu}(E_c)+\mathcal{M}^{0\nu}_{\alpha}.
\end{equation}
In this context we want to mention that in Ref. \cite{PhysRevC.88.064312} NMEs in closure method part of the above Eq. (\ref{eq:nmemixed}) was calculated in pure closure method \cite{PhysRevC.81.024321}.  Shell model code NushellX@MSU \cite{brown2014shell}, which was used in our calculation, do not give direct option to calculate required two body transition density (TBTD) \cite{PhysRevC.81.024321} for pure closure method. Hence, we calculate NMEs in closure method using Eq. (\ref{Eq:NMETNA}) in terms of TNA, which was first introduced in Ref. \cite{PhysRevLett.113.262501} and also used in Ref. \cite{PhysRevC.101.014307}.
One can expect a negligible difference in NMEs with closure and pure closure methods when TNA satisfies the Eq. (\ref{eq:tnap^2}) and (\ref{eq:tnan^2}).

Mixed methods has very good convergence property \cite{PhysRevC.88.064312}. Thus, this method is  particularly useful for calculating NMEs for higher mass region isotopes. Because of high convergence, NMEs calculated with few states of intermediate nucleus can give almost constant NMEs. 
\section{Results and Discussion}
\begin{table*}
\caption{\label{tab:nmet1} Nuclear matrix elements $M_F$, $M_GT$, $M_T$, $M_\nu$ for $0\nu\beta\beta$ of $^{48}$Ca, calculated with GXPF1A interaction in closure, running closure, running nonclosure and mixed methods for different SRC parametrization. $\langle E\rangle=7.72$ MeV was used for closure and running closure methods.}
\begin{ruledtabular}
\begin{tabular}{cccccc}
NME&SRC&Closure&Running closure&Running nonclosure&Mixed\\ \hline
$M_F$&None
     &-0.207&-0.206&-0.210&-0.211
\\
$M_F$&Miller-Spencer
&-0.141&-0.141&-0.143&-0.143

\\
$M_F$&CD-Bonn
   &-0.222&-0.221&-0.226&-0.227
\\
$M_F$&AV18
 &-0.204&-0.203&-0.207&-0.208
\\
   \\
$M_{GT}$&None
    &0.711&0.709&0.779&0.781
\\
$M_{GT}$&Miller-Spencer
  &0.492&0.490&0.553&0.555
\\
$M_{GT}$&CD-Bonn
  &0.738&0.736&0.810&0.812
\\
$M_{GT}$&AV18
&0.675&0.673&0.745&0.747
\\
   \\
$M_T$&None
 &-0.074&-0.072&-0.074&-0.076
\\
$M_T$&Miller-Spencer
&-0.076&-0.073&-0.075&-0.078
\\
$M_T$&CD-Bonn
&-0.076&-0.074&-0.076&-0.078
\\
$M_T$&AV18
&-0.077&-0.074&-0.076&-0.079
\\
  \\
$M_\nu$&None
&0.765&0.765&0.836&0.836
\\
$M_\nu$&Miller-Spencer
&0.504&0.505&0.566&0.565
\\
$M_\nu$&CD-Bonn
&0.799&0.799&0.874&0.874
\\
$M_\nu$&AV18
&0.725&0.725&0.798&0.798
\\
\end{tabular}
\end{ruledtabular}
\end{table*}
\begin{table*}
\caption{\label{tab:nmet2} Nuclear matrix elements $M_{\omega F}$, $M_{\omega GT}$, $M_{\omega T}$, $M_{\nu \omega}$ for $0\nu\beta\beta$ of $^{48}$Ca calculated with GXPF1A interaction in closure, running closure, running nonclosure and mixed methods for different SRC parametrization. $\langle E\rangle=7.72$ MeV was used for closure and running closure methods.}
\begin{ruledtabular}
\begin{tabular}{cccccc}
NME&SRC&Closure&Running closure&Running nonclosure&Mixed\\ \hline
$M_{\omega F}$&None
&-0.199&-0.198&-0.206&-0.207
\\
$M_{\omega F}$&Miller-Spencer
&-0.137&-0.136&-0.141&-0.142
\\
$M_{\omega F}$&CD-Bonn
&-0.212&-0.211&-0.220&-0.221
\\
$M_{\omega F}$&AV18
&-0.195&-0.194&-0.202&-0.203
\\
   \\
$M_{\omega GT}$&None
&0.66&0.659&0.766&0.767
\\
$M_{\omega GT}$&Miller-Spencer
&0.454&0.452&0.546&0.548
\\
$M_{\omega GT}$&CD-Bonn
&0.683&0.682&0.794&0.795
\\
$M_{\omega GT}$&AV18
&0.623&0.622&0.731&0.732
\\
   \\
$M_{\omega T}$&None
&-0.072&-0.069&-0.073&-0.076
\\
$M_{\omega T}$&Miller-Spencer
&-0.073&-0.070&-0.074&-0.077
\\
$M_{\omega T}$&CD-Bonn
&-0.074&-0.071&-0.075&-0.078
\\
$M_{\omega T}$&AV18
&-0.074&-0.071&-0.075&-0.078
\\
  \\
$M_{\nu \omega}$&None
 &0.712&0.712&0.821&0.821
\\
$M_{\nu \omega}$&Miller-Spencer
&0.466&0.467&0.559&0.558
\\
$M_{\nu \omega}$&CD-Bonn
&0.740&0.741&0.856&0.855
\\
$M_{\nu \omega}$&AV18
&0.670&0.671&0.781&0.780
\\
\end{tabular}
\end{ruledtabular}
\end{table*}
\begin{table*}
\caption{\label{tab:nmet3} Nuclear matrix elements $M_{q F}$, $M_{q GT}$, $M_{q T}$, $M_{1+}$, and $M_{2-}$ for $0\nu\beta\beta$ of $^{48}$Ca calculated with GXPF1A interaction in closure, running closure, running nonclosure and mixed methods for different SRC parametrization. $\langle E\rangle=7.72$ MeV was used for closure and running closure methods.}
\begin{ruledtabular}
\begin{tabular}{cccccc}
NME&SRC&Closure&Running closure&Running nonclosure&Mixed\\ \hline
$M_{q F}$&None
&-0.102&-0.102&-0.101&-0.101
\\
$M_{q F}$&Miller-Spencer
   &-0.082&-0.082&-0.080&-0.080
\\
$M_{q F}$&CD-Bonn
   &-0.123&-0.122&-0.121&-0.122
\\
$M_{q F}$&AV18
   &-0.118&-0.118&-0.117&-0.117
\\
   \\
$M_{q GT}$&None
     &3.243&3.246&3.317&3.314
\\
$M_{q GT}$&Miller-Spencer
   &2.681&2.684&2.751&2.748
\\
$M_{q GT}$&CD-Bonn
   &3.554&3.557&3.709&3.706
\\
$M_{q GT}$&AV18
   &3.423&3.426&3.502&3.499
\\
   \\
$M_{q T}$&None
  &-0.147&-0.140&-0.143&-0.150
\\
$M_{q T}$&Miller-Spencer
  &-0.150&-0.143&-0.146&-0.153
\\
$M_{q T}$&CD-Bonn
&-0.149&-0.142&-0.145&-0.153
\\
$M_{q T}$&AV18
&-0.150&-0.142&-0.146&-0.153
\\
  \\
$M_{1+}$&None
&3.937&3.898&3.989&4.028
\\
$M_{1+}$&Miller-Spencer
&3.430&3.389&3.480&3.521
\\
$M_{1+}$&CD-Bonn
&4.221&4.183&4.356&4.394
\\
$M_{1+}$&AV18
&4.101&4.061&4.158&4.198
\\
   \\
$M_{2-}$&None
&0.275&0.279&0.378&0.374
\\
$M_{2-}$&Miller-Spencer
&0.085&0.090&0.172&0.167
\\
$M_{2-}$&CD-Bonn
&0.271&0.276&0.372&0.367
\\
$M_{2-}$&AV18
&0.214&0.220&0.319&0.313
\\
\end{tabular}
\end{ruledtabular}
\end{table*}
\begin{figure*}
\centering
\includegraphics[trim=3cm 8.2cm 2.5cm 2cm,height=8.5cm,width=\linewidth]{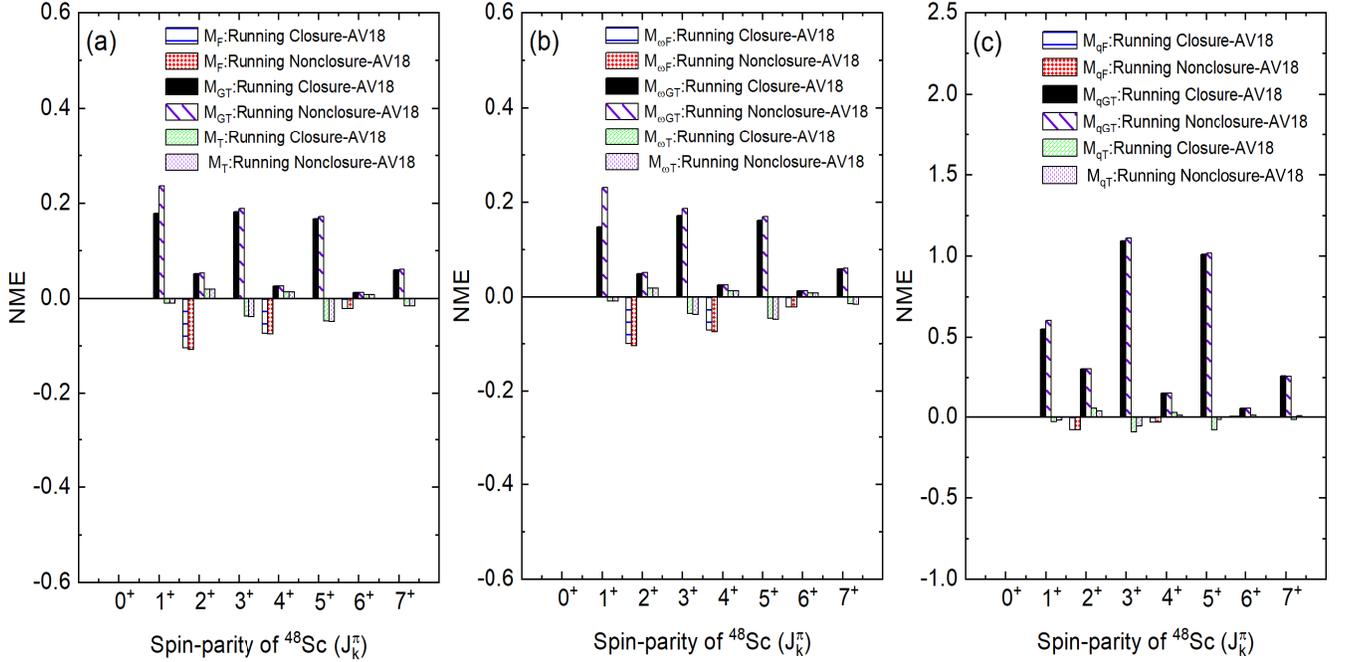}
\caption{\label{fig:NMEvsJk}(Color online) Contribution through different spin-parity of virtual intermediate states of $^{48}$Sc ($J_{k}^{\pi}$) in NMEs for $m_{\beta\beta}$ and $\lambda$ mechanisms of $0\nu\beta\beta$ of $^{48}$Ca. Here, comparison are shown for NMEs, calculated in running closure and running nonclosure methods with GXPF1A effective interaction for AV18 SRC parametrization. $\langle E\rangle=7.72$ MeV was used for running closure method.}
\end{figure*}
\begin{figure*}
\centering
\includegraphics[trim=3cm 8cm 2.5cm 2cm,height=8.2cm,width=\linewidth]{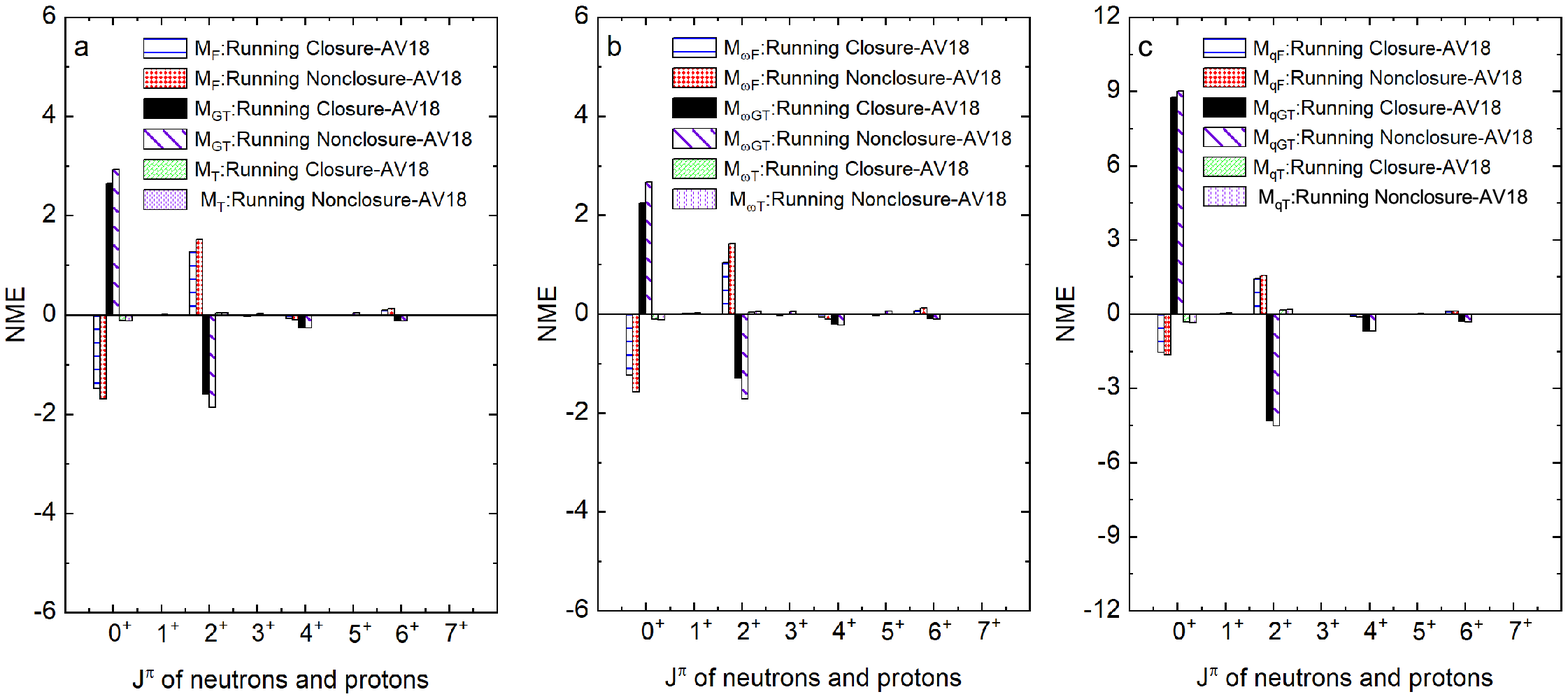}
\caption{\label{fig:NMEvsJ}(Color online) Contribution through different coupled spin-parity of two initial neutrons or two final created protons ($J^{\pi}$) in NMEs for $m_{\beta\beta}$ and $\lambda$ mechanisms of $0\nu\beta\beta$ of $^{48}$Ca. Here, comparison are shown for NMEs, calculated in running closure and running nonclosure methods with GXPF1A effective interaction for AV18 SRC parametrization. $\langle E\rangle=7.72$ MeV was used for running closure method.}
\end{figure*}
All necessary OBTD and TNA were calculated using shell-model code NushellX@MSU \cite{brown2014shell}. For calculating OBTD, first 150 states were considered for each allowed spin-parity ($J_{k}^{\pi}$) of virtual intermediate state $^{48}$Sc for $0\nu\beta\beta$ of $^{48}$Ca. TNA are calculated with $^{46}$Ca as an intermediate state for $0\nu\beta\beta$ of $^{48}$Ca. First 100 states were considered for each allowed spin-parity of $^{46}$Ca ($J_{m}^{\pi}$). Anti-symmetric non reduced two-body matrix elements for closure method and reduced non anti-symmetric two-body matrix elements for running closure and running nonclosure methods are calculated by program written by us. Calculated NMEs $M_F$, $M_{GT}$, $M_T$, $M_\nu$ are given in Table \ref{tab:nmet1}. The $M_{\omega F}$, $M_{\omega GT}$, $M_{\omega T}$, $M_{\nu \omega}$ type NMEs are presented in Table \ref{tab:nmet2} and the $M_{q F}$, $M_{q GT}$, $M_{q T}$, $M_{\nu q}$, $M_{1+}$, $M_{2-}$ type NMEs are given in Table \ref{tab:nmet3}. 

From Table \ref{tab:nmet1}, it is found for different SRC parametrization that the $M_{F}$ type NMEs calculated in the running closure method are near to NMEs in the closure method. In this case, NMEs in running nonclosure method are about 1-2\% larger in magnitude than the corresponding NMEs in the running closure method. The NMEs in the mixed method are also close to NMEs in running nonclosure method. The 

$M_{GT}$ type NMEs calculated in the running closure method are near to NMEs in the closure method. In this case, NMEs in running nonclosure method are about 10-13\% larger than the corresponding NMEs in the running closure method. The NMEs in the mixed method are also close to NMEs in running nonclosure method. 

For different SRC parametrization, $M_{T}$ type NMEs calculated in the running closure method are about 3-4\% smaller in magnitude than the corresponding NMEs in the closure method. In this case, the NMEs in running nonclosure method are about 3\% larger in magnitude than the corresponding NMEs in running closure method, and the NMEs in the mixed method are about 3-4\% larger than the corresponding NMEs in running nonclosure method.

The $M_{\nu}$ type NMEs calculated in running closure method are near to NMEs in closure method for different SRC parametrization. In this case, NMEs in running nonclosure method is about 9-12\% larger than the corresponding NMEs in running closure method. The NMEs in mixed method in this case is also close to the NMEs in running nonclosure method. The dominating enhancement in nonclosure NMEs are seen through Gamow-Teller type NMEs.
\begin{figure*}[t]
\centering
\includegraphics[trim=2.5cm 2cm 3.5cm 2cm,width=\linewidth]{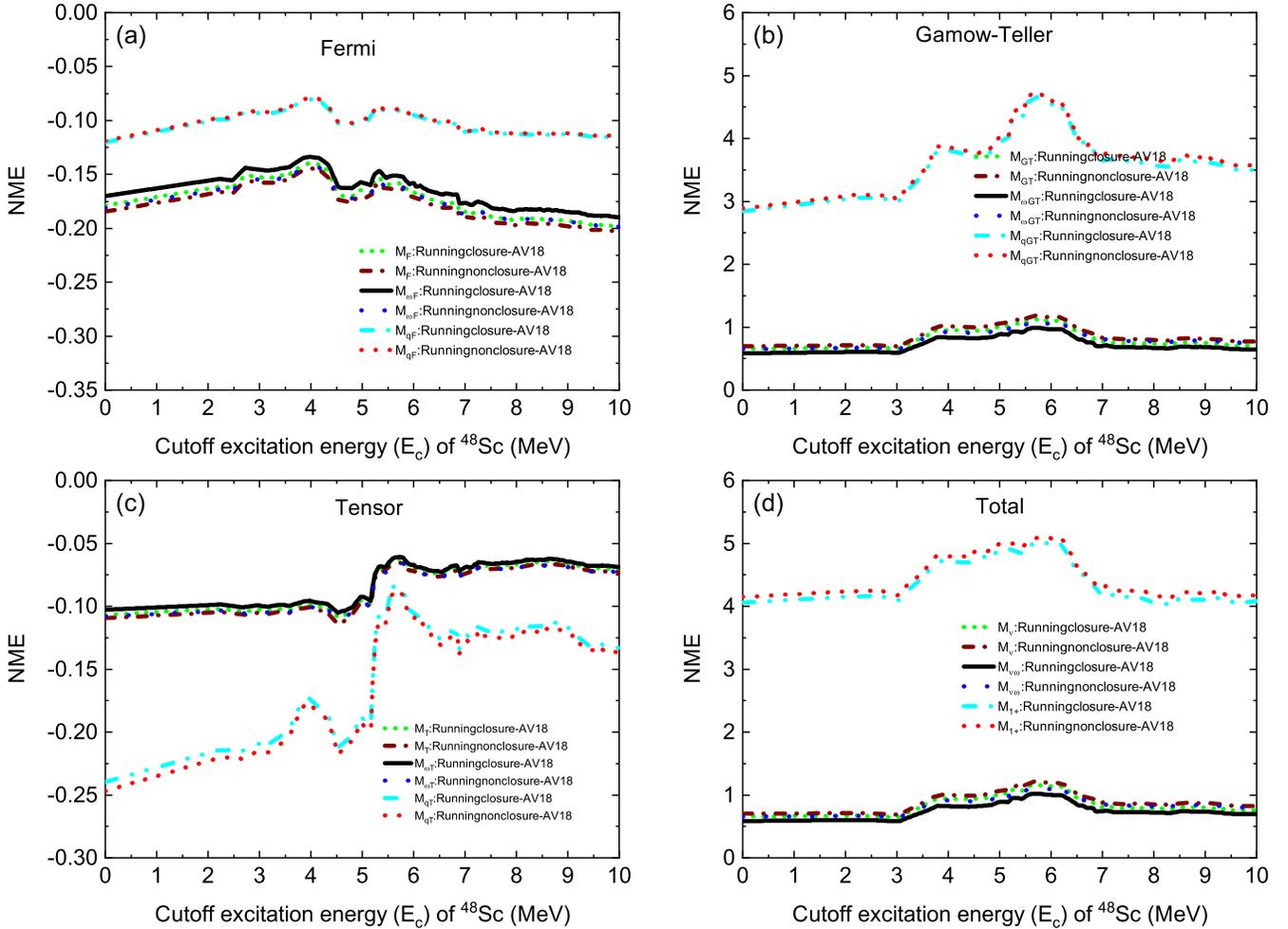}
\caption{\label{fig:nmevsek}(Color online) Variation of (a) Fermi (b) Gamow-Teller (c) tensor and (d) total NMEs for $0\nu\beta\beta$ ($m_{\beta\beta}$ and $\lambda$ mechanisms) of $^{48}$Ca with cutoff excitation energy ($E_c$) of states of virtual intermediate nucleus $^{48}$Sc. NMEs are calculated with total GXPF1A interaction for AV18 SRC parametrization in running closure and running nonclosure methods. For running closure method, closure energy $\langle E\rangle $=7.72 MeV was used.}
\end{figure*}

From Table \ref{tab:nmet2}, it is found for different SRC parametrization that the $M_{\omega F}$ type NMEs calculated running nonclosure method are about 4\% larger in magnitude than the corresponding NMEs in running closure method.

The $M_{\omega GT}$ type NMEs calculated in running nonclosure method are about 16-21\% larger than the corresponding NMEs in the running closure method. The $M_{\omega T}$ type NMEs calculated in the running closure method are about 4\% smaller in magnitude than the NMEs in the closure method. In this case, the NMEs in running nonclosure method are about 6\% larger in magnitude than the corresponding NMEs in running closure method, and the NMEs in mixed methods are about 4\% larger than the corresponding NMEs in running nonclosure method for different SRC parametrization.

It is found for different SRC parametrization that there are very small differences of the $M_{\nu\omega}$ type NMEs calculated in running closure and closure methods, which lead to similar NMEs in running nonclosure and mixed methods. In this case, NMEs in running nonclosure method are about 15-20\% larger than the corresponding NMEs in the running closure method. Most of the enhancements come through $M_{\omega GT}$ type NMEs. 

From Table \ref{tab:nmet3}, it is found for different SRC parametrization that the $M_{qF}$ type NMEs calculated in running nonclosure method are about 1-2\% smaller in magnitude than the corresponding NMEs in running closure method. The $M_{qGT}$ type NMEs calculated in running nonclosure method are about 2-4\% larger than the corresponding NMEs in the running closure method. The $M_{qT}$ type NMEs calculated in the running closure method are about 5\% smaller in magnitude than the corresponding NMEs in the closure method. In this case, the NMEs in running nonclosure method are about 2-3\% larger than the corresponding NMEs in the running closure method, and the NMEs in the mixed method are about 5\% larger than the corresponding NMEs in running nonclosure method for different SRC parametrization. Both $M_{qF}$ and $M_{qGT}$ type have similar values in closure and running closure methods, which lead to similar values of those NMEs in running nonclosure and mixed methods. 

The $M_{1+}$ type NMEs calculated in the running closure method are 1\% smaller than the corresponding NMEs in closure method for different SRC parametrization. This total NMEs in running nonclosure method is about 2-4\% larger than the corresponding NMEs in the running closure method. In this case, the NMEs in the mixed method are about 1\% larger than the corresponding NMEs in the running nonclosure method. 

The $M_{2-}$ type NMEs calculated in the running closure method are about 2-6\% larger than the corresponding NMEs in the closure method for different SRC parametrization. In this case, the $M_{2-}$ type NMEs in running nonclosure method are about  35-91\% larger than the corresponding NMEs in the running closure method. The NMEs in the mixed method are about 1-3\% smaller than the corresponding NMEs in running nonclosure method. 

From, Table \ref{tab:nmet1},\ref{tab:nmet2},\ref{tab:nmet3}, it was found that  $M_{\nu\omega}$ NMEs are about 7-8\%, 7-8\%, 1-2\%, 1-2\% smaller than $M_{\nu}$ type NMEs in closure, running closure, running nonclosure, and mixed methods, respectively for different SRC parametrization. 
 The $M_{1+}$ type NMEs are about 453-636\%, 447-625\%, 385-522\%, 390-530\% larger than $M_{\nu\omega}$ type NMEs in closure, running closure, running nonclosure, and mixed methods, respectively for different SRC parametrization. 
 
 This increment of  $M_{1+}$ type of NME is surprisingly high, which is coming through the very large value of $M_{qGT}$ type NME as compared to $M_{GT}$, and $M_{\omega GT}$ type NMEs. 
 To justify the large value of $M_{qGT}$ type NME, we have carefully checked our calculations and found that large value $M_{qGT}$ are coming through the new revisited expression of the nucleon currents of Ref. \cite{vsimkovic2017lambda} which includes higher-order term (isoscalar) of the nucleon currents. In our calculation, Eq. (\ref{Eq:hqgt}) was used to calculate $M_{qGT}$ type NME using the revisited formalism of nucleon currents of Refs. \cite{vsimkovic2017lambda,PhysRevC.92.055502}.  
An old equivalent expression of Eq. (\ref{Eq:hqgt}) was also found in  Ref. \cite{PhysRevC.98.035502}, which one can write using Eq. (A2c) and Eq. (A4b) of Ref. \cite{PhysRevC.98.035502} as
\begin{equation}
\label{eq:couplingold}
   f_{qGT}(q,r)=\frac{1}{(1+\frac{q^{2}}{\Lambda_{A}^{2}})^{4}}qrj_{1}(qr).
\end{equation}
Using Eq. (\ref{eq:couplingold}), which does not include the higher-order terms of nucleon currents, $M_{qGT}$ type NME for $^{48}$Ca was reported to be 0.709 in Table XVI of Ref. \cite{PhysRevC.98.035502} for CD-Bonn type SRC parametrization with closure energy $\langle E\rangle=0.5$ MeV. We have also performed the same calculation for $M_{qGT}$ type NMEs in closure method for CD-Bonn SRC parametrization with $\langle E\rangle=0.5$ MeV by using old factor of Eq. (\ref{eq:couplingold}) instead of new factor of Eq. (\ref{Eq:hqgt}). In this case the value of $M_{qGT}$ was found to be 0.708, which is almost same as the value reported in Ref. \cite{PhysRevC.98.035502}. Thus we infer that the large value of $M_{qGT}$ type NME is coming through the revisited nucleon currents of Ref. \cite{vsimkovic2017lambda} which includes the higher-order term (isoscalar term) of the nucleon currents. 
\begin{figure*}[t]
\centering
\includegraphics[trim=2.5cm 2cm 3.5cm 2cm,width=\linewidth]{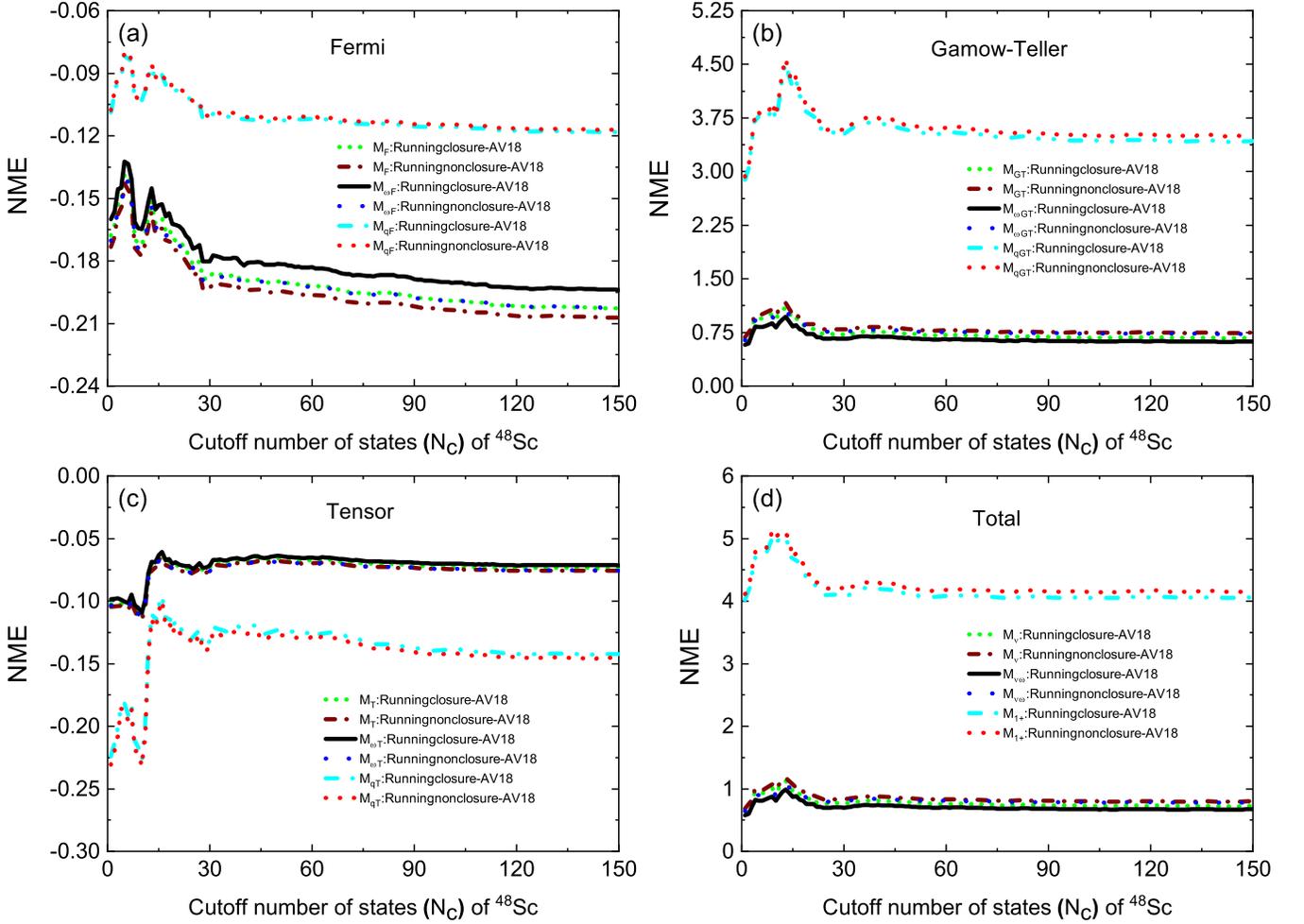}
\caption{\label{fig:nmevsnk}(Color online) Variation of (a) Fermi (b) Gamow-Teller (c) tensor and (d) total NMEs for $0\nu\beta\beta$ ($m_{\beta\beta}$ and $\lambda$ mechanisms) of $^{48}$Ca with cutoff number of states ($N_c$) of virtual intermediate nucleus $^{48}$Sc. NMEs are calculated with total GXPF1A interaction for AV18 SRC parametrization in running closure and running nonclosure methods. For running closure method, closure energy $\langle E\rangle $=7.72 MeV was used.}
\end{figure*}

To examine the contribution of each allowed spin-parity ($J_{k}^{\pi}$) of virtual intermediate states of $^{48}$Sc, using Eq. (\ref{eq:rcpartial})  one can write NMEs as function of $J_{k}^{\pi}$ in running closure method as
\begin{eqnarray}
\mathcal{M}_{\alpha}^{0\nu}(E_c,J_k)=\sum_{J,E_{k}^{*}\leqslant E_c}\mathcal{M}_{\alpha}^{0\nu}(J_{k},J,E_{k}^{*})
\end{eqnarray}
 and using Eq. (\ref{eq:rncpartial}), one can write in running nonclosure method as
\begin{eqnarray}
{M}_{\alpha}^{0\nu}(E_c,J_k)=\sum_{J,E_{k}^{*}\leqslant E_c}{M}_{\alpha}^{0\nu}(J_{k},J,E_{k}^{*})
\end{eqnarray}
Dependence of Fermi, Gamow-Teller, and tensor NMEs with different $J_k^{\pi}$ of $^{48}$Sc are shown in Fig.~\ref{fig:NMEvsJk}. Here NMEs are calculated in running closure and running nonclosure methods with GXPF1A effective interaction for AV18 SRC parametrization. It is found that for all Fermi type NMEs, contribution through each $J_{k}^{\pi}$ is negative and for Gamow-Teller all contribution is positive. Contribution in tensor NMEs comes in opposite phase for different $J_{k}^{\pi}$ reducing the total tensor NMEs. 

Further, it is found that the dominating contribution comes from 2$^{+}$ state for $M_{F}$ and $M_{\omega F}$ type NMEs and with a small contribution from 4$^{+}$  and 6$^{+}$  states. Contributions through 0$^{+}$  and odd-$J_k^{\pi}$ states are negligible. For $M_{qF}$ type NME, the dominating contribution comes through 2$^{+}$ and 4$^{+}$ states. For all Fermi type NME, enhancements in running nonclosure method are very small as compared to running closure method. 

For $M_{GT}$, $M_{\omega GT}$, and $M_{qGT}$   type NMEs, all $J_{k}^{\pi}$ contributes significantly except 0$^{+}$. Dominating contribution come through 1$^{+}$, 3$^{+}$, and 5$^{+}$ states for $M_{GT}$, $M_{\omega GT}$ type NMEs, whereas, for $M_{qGT}$ type NMEs, order of most dominating contributions are through 3$^{+}$, 5$^{+}$, and 1$^{+}$ states. Significant enhancement in NMEs with running nonclosure method was found through 1$^{+}$ states as compared to NMEs in the running closure method.    

For $M_{T}$, $M_{\omega T}$ type NMEs, all negative contributions come through 1$^{+}$, 3$^{+}$, 5$^{+}$, and 7$^{+}$ states with contribution from 3$^{+}$, and 5$^{+}$ states being dominating. Contributions through 2$^{+}$, 4$^{+}$, and 6$^{+}$ are all positive. For $M_{qT}$ type NMEs bulk of negative contributions comes through 3$^{+}$ state with additional negative contribution comes through 1$^{+}$ and 5$^{+}$ states. Dominating positive contribution comes through 2$^{+}$ states with a very small positive contribution thorough 4$^{+}$ state. A similar pattern of variation of different types of NMEs with $J_{k}^{\pi}$ are seen with other types of SRC parametrization. 

We have also checked the variation of NMEs with coupled spin-parity ($J^\pi$) of two decaying neutrons and two created protons in the decay. One can write using Eq.~(\ref{eq:rcpartial}) and (\ref{eq:rncpartial}) NMEs as function of $J^\pi$ in running closure method as
\begin{eqnarray}
\mathcal{M}_{\alpha}^{0\nu}(E_c,J)=\sum_{J_k,E_{k}^{*}\leqslant E_c}\mathcal{M}_{\alpha}^{0\nu}(J_{k},J,E_{k}^{*})
\end{eqnarray}
and in running nonclosure method as
\begin{eqnarray}
{M}_{\alpha}^{0\nu}(E_c,J)=\sum_{J_k,E_{k}^{*}\leqslant E_c}{M}_{\alpha}^{0\nu}(J_{k},J,E_{k}^{*})
\end{eqnarray}
Contributions of NMEs through different $J^\pi$ is shown in Fig.~\ref{fig:NMEvsJ}. Here NMEs are calculated in running closure and running nonclosure method for AV18 SRC parmaetrization. 

For all types of NMEs, the most dominating contribution comes from 0$^{+}$ states and 2$^{+}$ states. Also, the contribution through 0$^{+}$ and 2$^{+}$ states comes in opposite sign reducing the total NMEs. The small contribution comes through 4$^{+}$ and 6$^{+}$ states with almost negligible contributions from odd-$J^\pi$ states. Pairing effect is responsible for dominating even-$J^\pi$ contributions \cite{PhysRevLett.113.262501}. Significant enhancement was found in NMEs calculated with running nonclosure method through both  0$^{+}$  and 2$^{+}$ states as compared to NMEs in running closure method. 
\begin{figure}
\centering
\includegraphics[trim=2cm 1cm 2cm 2cm,width=\linewidth]{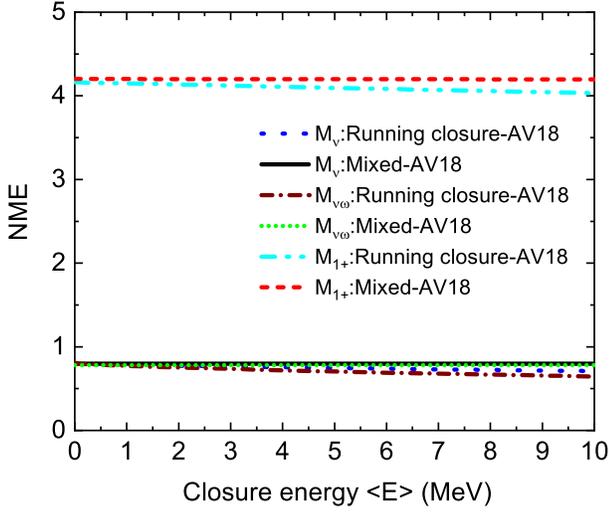}
\caption{\label{fig:nmevsclosure}(Color online) Dependence of the total NMEs for $0\nu\beta\beta$ ($\lambda$ and $m_{\beta\beta}$ mechanisms) of $^{48}$Ca with closure energy $\langle E\rangle$, calculated with total GXPF1A interaction for AV18 SRC parmaetrization in running closure and mixed methods.}
\end{figure}

Dependence of NMEs with cutoff excitation energy ($E_c$) of the intermediate nucleus $^{48}$Sc was examined. One can write using Eq. (\ref{eq:rcpartial}), NMEs as function $E_c$ in running closure method as
\begin{eqnarray}
\mathcal{M}_{\alpha}^{0\nu}(E_c)=\sum_{J_k,J,E_{k}^{*}\leqslant E_c}\mathcal{M}_{\alpha}^{0\nu}(J_{k},J,E_{k}^{*})
\end{eqnarray}
and using Eq. (\ref{eq:rncpartial}) NMEs in running nonclosure method as
\begin{eqnarray}
{M}_{\alpha}^{0\nu}(E_c)=\sum_{J_k,J,E_{k}^{*}\leqslant E_c}{M}_{\alpha}^{0\nu}(J_{k},J,E_{k}^{*}).
\end{eqnarray}
Variation of Fermi, Gamow-Teller, tensor, and total NMEs with cutoff excitation energy ($E_c$) of $^{48}$Sc is shown in Fig. \ref{fig:nmevsek}. Here, we have shown the dependence of NMEs for $E_c$=0 to 10 MeV. The NMEs are calculated in running closure and running nonclosure methods for AV18 type SRC parametrization. It is observed that first few low lying states for each $J_k^{\pi}$ mostly contributes constructively and destructively. After about $E_c$=7 MeV, NMEs attains mostly stable value. At large enough value of $E_c$ NMEs becomes constant. This is because NMEs are less sensitive with excitation energy of $^{48}$Sc as the large neutrino momentum $q$, which is about $\sim$ 100-200 MeV sitting in the denominator of the neutrino potential in Eq. (\ref{eq:npnc}). A similar variation of NMEs with $E_c$ are found for other SRC parametrization for different types of NMEs. 

It is also possible to set a cutoff on the number of states ($N_c$) for each allowed $J_{k}^{\pi}$ of $^{48}$Sc to calculate the NMEs. One can write the NMEs as function of cutoff number of states ($N_c$) of $^{48}$Sc in running closure method as
\begin{eqnarray}
\mathcal{M}_{\alpha}^{0\nu}(N_c)=\sum_{J_k,J,N_k\leqslant N_c}\mathcal{M}_{\alpha}^{0\nu}(J_{k},J,N_k),
\end{eqnarray}
and in running nonclosure method as
\begin{eqnarray}
{M}_{\alpha}^{0\nu}(N_c)=\sum_{J_k,J,N_k\leqslant N_c} {M}_{\alpha}^{0\nu}(J_{k},J,N_k),
\end{eqnarray}
where $\mathcal{M}_{\alpha}^{0\nu}(J_{k},J,N_k)$ and ${M}_{\alpha}^{0\nu}(J_{k},J,N_k)$ is same as Eq. (\ref{eq:rcpartial}), and (\ref{eq:rncpartial}), respectively. 
Dependence of different Fermi, Gamow-Teller, tensor, and total NMEs with $N_c$ is shown in Fig.~\ref{fig:nmevsnk}. Here, NMEs are calculated with total GXPF1A interaction in running closure and running nonclosure methods for AV18 SRC parametrization. The dependence shows that the first few low lying states mostly contribute constructively and destructively. After $N_c$=75, different types of NMEs attain a stable value. At a large value of $N_c$, NMEs becomes mostly constant. In our calculation, we have considered $N_c$=150 for each allowed $J_k^{\pi}$ of $^{48}$Sc, which gives NMEs with negligible uncertainty. A similar dependence of NMEs with $N_c$ are seen for other SRC parametrizations. 

In the end, we have shown the variation of different types of total NMEs in running closure and mixed methods with closure energy $\langle E\rangle$ in Fig. \ref{fig:nmevsclosure}. NMEs shown here are calculated with total GXPF1A interaction for AV18 type SRC parametrization.
It is observed that for changing $\langle E\rangle$=0 to 10 MeV, there are about 11\% decrements of $M_{\nu}$ type NMEs in running closure method. In this case, $M_{\nu\omega}$ type NMEs decrease by about 19\%, and $M_{1+}$ type NMEs decrease by about 3\% in running closure method. In all these cases, we found negligible changes in NMEs in the mixed method. A similar pattern of variation of NMEs with $\langle E\rangle$ are found with other SRC parametrizations. 

\section{\label{sec:VI}Summary and Conclusions}
Different types of NMEs for $m_{\beta\beta}$ and $\lambda$ mechanisms of $0\nu\beta\beta$, which has origin in the left-right symmetric model with right-handed gauge boson at TeV scale, were calculated in ISM for one of the $0\nu\beta\beta$ decay candidate, $^{48}$Ca. The GXPF1A effective interaction of $pf$-shell was used to calculate NMEs in both closure, and nonclosure approaches. The standard effects of FNS and the revisited higher-order terms such as isoscalar and weak magnetism terms of nucleon currents, which were exploited in Refs. \cite{PhysRevC.92.055502,vsimkovic2017lambda}, were also included the present work. The short-range nature of nucleon-nucleon interactions was also taken care of in Miller-Spencer, CD-Bonn, and AV18 SRC parametrizations. Detailed comparative results were presented for closure versus nonclosure approaches using four different methods: closure, running closure, running nonclosure, and mixed methods. The significant enhancements of $M_{qGT}$ and $M_{qT}$ type NMEs were found for the inclusion of new isoscalar term of nucleon currents.  

Results show that $M_{\nu}$ type NMEs in the nonclosure approach is about 9-12\% larger than the corresponding NMEs in the closure approach. For different SRC parametrization, the $M_{\nu\omega}$ type NMEs in the nonclosure approach is about 15-20\% larger than the corresponding NMEs in the closure approach, and $M_{1+}$ type NMEs in nonclosure approach is about 2-4\% larger than the corresponding NMEs in closure approach. The $M_{qGT}$ type NME is found to be much larger than the $M_{GT}$, and $M_{\omega GT}$ type NMEs, which is coming for the inclusion isoscalar term of the nucleon currents for $M_{qGT}$ NMEs. 

We have examined the contribution of each spin-parity ($J_k^{\pi}$) of the intermediate nucleus $^{48}$Sc in NMEs running closure and running nonclosure methods. It is observed that
$2^{+}$ and $4^{+}$ states have dominant contribution in $M_{F}$, $M_{\omega F}$, and $M_{qF}$ type NMEs. For $M_{GT}$, $M_{\omega GT}$, and $M_{qGT}$ type NMEs, $1^{+}$, $3^{+}$, and $5^{+}$ states contribute the most. In this case, the dominating enhancements in NMEs with nonclosure approach are seen through $1^{+}$ states as compared to the corresponding NMEs in closure approach. Contributions of all $J_{k}^{\pi}$ are negative for the Fermi type NMEs and positive for the Gamow-Teller type NMEs. For $M_{T}$, $M_{\omega T}$ type NMEs all contributions are negative for 
$1^{+}$, $3^{+}$,  $5^{+}$, and $7^{+}$ states and positive for $2^{+}$, $4^{+}$,and $6^{+}$ states. For $M_{qT}$ type NME, dominating negative contribution comes through $3^{+}$state and positive contribution comes through $2^{+}$ state. 

We have also checked the dependence of the NMEs calculated in running nonclsoure and running closure method with coupled spin-parity of two initial neutrons or final created protons ($J^\pi$) of the decay. It is found that dominating contribution comes through 0$^{+}$ and 2$^{+}$ states with their phase being opposite, which reduce the total NMEs. Enhancements in NMEs with nonclosure approach are seen through both 0$^{+}$ and 2$^{+}$ states. 

Dependence of different types of NMEs 
with the cutoff excitation energy ($E_c$) and the cutoff number of states ($N_c$) of intermediate nucleus ($^{48}$Sc) are also explored. It is found that only the first few low lying states contribute constructively and destructively in NMEs, and at large $N_c$ and $E_c$, NMEs become almost constant. This is because of the large value of neutrino momentum $q$ ($\sim$ 100-200 MeV), whereas relevant excitation energy is only of the order of 10 MeV. In our case, we have considered $N_c=$150 for each $J_{k}^{\pi}$ of $^{48}$Sc with uncertainty being very small.

Variation of NMEs in running closure and mixed methods with closure energy $\langle E\rangle$ are also studied. For changing $\langle E\rangle$=0 to 10 MeV, there is about 11\% decrements of $M_{\nu}$ type NME in running closure method. In this case $M_{\nu\omega}$ type NME decreases by about 19\%  and $M_{1+}$ type NME decreases by about 2.5\% in running closure method. In all these cases, there are negligible change in NMEs in mixed method.

\begin{acknowledgments}
S.S. thanks Ministry of Human Resource Development (MHRD), Government of India for the financial assistance towards Ph.D. progress. Y.I. 
acknowledges the support from JSPS KAKENHI Grant No.17K05440.  
\end{acknowledgments}
\nocite{}
\bibliography{main}
\end{document}